\pdfoutput=1

\documentclass[11pt]{article}

\usepackage[final]{acl}

\usepackage{times}
\usepackage{latexsym}

\usepackage[T1]{fontenc}

\usepackage[utf8]{inputenc}

\usepackage{microtype}

\usepackage{inconsolata}

\usepackage{graphicx}

\usepackage{multirow}
\usepackage{enumitem}
\usepackage{booktabs}
\usepackage{array}
\usepackage{makecell}
\usepackage{todonotes}
\usepackage{layouts}
\usepackage{xcolor}
\usepackage{adjustbox}

\newcolumntype{L}{>{\centering\arraybackslash}m{1.2cm}}
\newcolumntype{P}{>{\centering\arraybackslash}p{1.2cm}}

\title{CLIRudit: Cross-Lingual Information Retrieval of Scientific Documents}

\author{
 \textbf{Francisco Valentini\textsuperscript{1,2}
 \thanks{Research conducted during a stay at the École de bibliothéconomie et des sciences de l'information, Université de Montréal, Canada.}}, 
 \textbf{Diego Kozlowski\textsuperscript{2}}, 
 \textbf{Vincent Larivi{è}re\textsuperscript{2}}
\\
\\
 \textsuperscript{1}CONICET-Universidad de Buenos Aires.\\Instituto de Ciencias de la Computación (ICC). Buenos Aires, Argentina\\
 \textsuperscript{2}École de bibliothéconomie et des sciences de l'information.\\Université de Montréal. Montréal, Canada\\
   \small{\texttt{fvalentini@dc.uba.ar},}
   \small{\texttt{diego.kozlowski@umontreal.ca,}}
   \small{\texttt{vincent.lariviere@umontreal.ca}}
}

\begin{document}
\maketitle
\begin{abstract}
    Cross-lingual information retrieval (CLIR) helps users find documents in languages different from their queries.
    This is especially important in academic search, where key research is often published in non-English languages. 
    We present CLIRudit, a novel English-French academic retrieval dataset built from Érudit, a Canadian publishing platform. 
    Using multilingual metadata, we pair English author-written keywords as queries with non-English abstracts as target documents, a method that can be applied to other languages and repositories.   
    We benchmark various first-stage sparse and dense retrievers, with and without machine translation.
    We find that dense embeddings without translation perform nearly as well as systems using machine translation, that translating documents is generally more effective than translating queries, and that sparse retrievers with document translation remain competitive while offering greater efficiency. 
    Along with releasing the first English-French academic retrieval dataset, we provide a reproducible benchmarking method to improve access to non-English scholarly content.

\end{abstract}

\section{Introduction} \label{sec:introduction}

\begin{figure}
  \centering
  \includegraphics[width=1\linewidth]{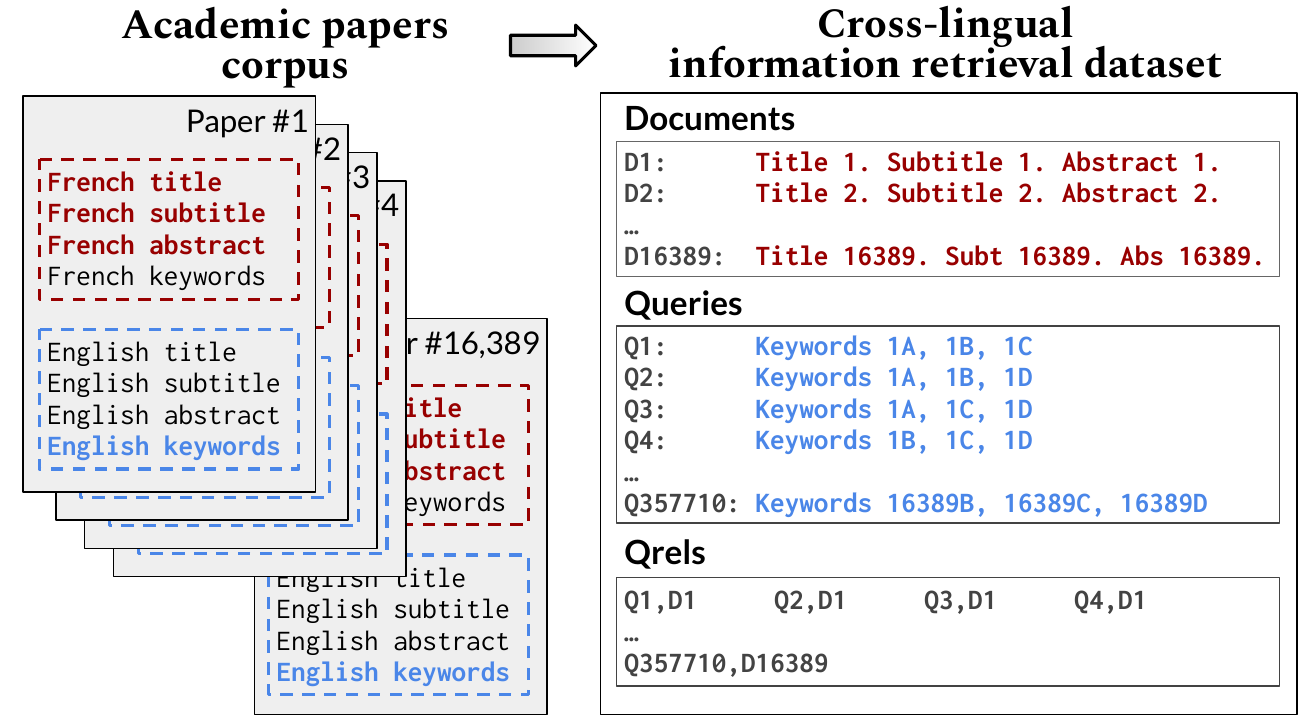}
  \caption{
    The CLIRudit dataset.
    We use articles with abstracts and keywords in both French and English.
    English keywords form the queries, with relevance judged by their presence in each article.
    Documents consist of the French title, subtitle, and abstract.
  }
  \label{fig:main_figure}
\end{figure}

Cross-lingual information retrieval (CLIR) helps users find documents written in languages different from their search queries.
This removes the need for proficiency in multiple languages and makes it easier to access valuable information that might otherwise be missed because of language barriers.

CLIR is especially important for academic research.
While English is the main language for scientific communication, important work often exists in other languages, particularly in certain fields and historical contexts \citep{polonen2020helsinki,beigel2022batalla,khanna2022recalibrating}.
Researchers may overlook key work if they cannot search across languages, especially if they're unfamiliar with technical terms.
English is also often used due to the expectation of finding more results, reinforcing bias against documents in other languages. 

Modern information retrieval (IR) systems often use bi-encoder architectures for first-stage retrieval, separately encoding documents and queries as dense embeddings \citep{devlin2019bert, karpukhin2020dense, xiong2021approximate}. 
Multilingual extensions of these methods have been effective in general-domain CLIR \citep{ artetxe2019massively, conneau2020unsupervised, anastasopoulos2020should, asai2021xor, nair2022transfer, zhang2023toward}.
Another common approach is to use machine translation (MT) to convert queries or documents to the same language before searching \citep{galuscakova2022clir,lin2022simple,huang2023improving,lawrie2024overview}.

Technical texts often use specialized vocabulary and styles that present challenges for MT and multilingual embeddings \citep{lawrie2024overview, litschko2025cross}. 
However, research on CLIR in technical domains is limited \citep{xu2016domain, zavorin2020corpora}, and studies focusing specifically on academic content are even scarcer, typically relying on small, curated datasets \citep{lawrie2024overview}.
As a result, the effectiveness of CLIR methods for academic retrieval remains underexplored. 


We address this gap by introducing a dataset for cross-lingual academic search and benchmarking first-stage retrieval methods.
Our contributions are:


\begin{itemize}[wide, itemindent=\labelsep]

  \item A new method for creating academic CLIR datasets using multilingual metadata. 
  We use English keywords as queries and non-English abstracts as documents, allowing evaluation of IR methods on retrieving original-language documents based on author-provided English keywords.
  This method can be applied to other academic databases and language pairs.

  \item The release of CLIRudit, a dataset based on Érudit, a Quebec-based non-profit publishing platform (Fig. \ref{fig:main_figure}).\footnote{\url{https://hf.co/datasets/ftvalentini/clirudit}} 
  To our knowledge, this is the first dataset for English-French academic retrieval.

  \item A thorough empirical comparison of first-stage CLIR methods, including query and document translation, and state-of-the-art dense and sparse retrievers.

  \item Practical insights to improve the discoverability of non-English scholarly content, which is especially relevant for academic publishing platforms.

\end{itemize}

Our results show that dense embeddings without translation perform nearly as well as those using MT.
Document translation generally improves retrieval more than query translation. 
While sparse retrievers combined with document translation may not surpass the best dense multilingual methods, they remain competitive and offer advantages in search speed and indexing efficiency.

\section{Related work} \label{sec:related_work}

This section reviews relevant research on academic CLIR, focusing on first-stage retrieval methods, datasets, and bilingual academic corpora.

\subsection{Cross-lingual retrieval} \label{sec:clir}

\citet{lin2022simple} proposed a conceptual framework for CLIR, outlining three main strategies for first-stage retrieval:
\textbf{document translation} (DT), translating documents into the query language; \textbf{query translation} (QT), translating queries into the document language; and \textbf{language-independent representations}, encoding queries and documents into a shared vector space for direct retrieval.
Since we focus on single-stage retrieval, we do not address later steps of a retrieval pipeline, such as re-ranking or results fusion.

Translation-based methods have been widely used and generally effective, although their success has varied across domains and language pairs.
DT combined with neural ranking has shown strong performance in general-domain tasks \citep{lin2022simple, lawrie2023neural, lassance2023naverloo}, often outperforming QT, which struggles with short, ambiguous queries and limited training data \citep{galuscakova2022clir}.
However, DT is not a clear winner, with QT performing better in domains like healthcare \citep{saleh2020document} and in high-resource languages \citep{huang2023improving}.  

Alternative approaches like probabilistic structured queries (PSQ) generate multiple plausible translations per term using alignment models, offering more flexibility than standard machine translation \citep{darwish2003probabilistic, yang2024efficiency}.

Early studies found a strong link between translation quality and retrieval effectiveness \citep{zhu2006effect}, but later work found that better MT doesn't always improve retrieval, particularly in specialized domains \citep{pecina2014adaptation}.
Recent research suggests a weak positive correlation \citep{bonifacio2022mmarco} with diminishing returns beyond a certain MT quality level \citep{zhang2022machine}.

Multilingual bi-encoders avoid MT entirely by using multilingual pretrained models \citep{jiang2020cross, bonifacio2022mmarco, nair2022transfer, nair2023blade}.
These methods can reduce indexing costs but often perform worse than MT-based retrieval, with QT or DT followed by monolingual retrieval frequently achieving better first-stage results \citep{litschko2019evaluating, asai2021xor, lin2022simple, nair2023blade, lawrie2023neural}.


Recent methods like translate-train \citep{nair2022transfer} and translate-distill \citep{yang2024translate} integrate MT into training, allowing bi-encoders to jointly learn retrieval and translation; unlike translate-test methods like DT and QT, which translate only at test-time.
Translate-distill further uses distillation from cross-encoders, achieving strong results across multiple languages.
Additionally, large decoder-only language models (LLMs) have been adapted as bi-encoders for dense retrieval  \citep{lee2024nv,li2025making}.


\subsection{CLIR datasets} \label{sec:datasets}

Well-documented and diverse datasets are crucial for advancing CLIR because they enable training and evaluation across languages and domains.

Shared evaluation initiatives like TREC \citep{voorhees2005trec} and CLEF \citep{chen2002cross} provide manually curated test collections with human-generated queries and relevance judgments gathered by pooling top-ranked results.
NeuCLIR (TREC 2022) focuses on neural CLIR, alongside other datasets such as BETTER \citep{soboroff2023better} and HC4 \citep{lawrie2022hc4}.
While these collections are usually carefully designed, they are typically small, often with fewer than 1,000 queries. \citet{galuscakova2022clir} provide a comprehensive survey of such resources.

Sentence-level retrieval datasets are also common, such as BUCC, Tatoeba \citep{siddhant2020xtreme}, and STS17/STS22 \citep{cer2017semeval, chen2022semeval}, which focus on matching similar sentences across languages.

To address scale limitations, recent work has explored automatic dataset creation.
For example, \citet{mayfield2023synthetic} used LLMs to generate English queries from target-language documents.
Wikipedia's multilingual, structured content has also been used for automatic dataset creation, as seen in MuSeCLIR \citep{li2022museclir}, MKQA \citep{longpre2021mkqa}, WikiCLIR \citep{sasaki2018cross}, CLIRMatrix \citep{sun2020clirmatrix}, and AfriCLIRMatrix \citep{ogundepo2022africlirmatrix}.

\subsection{Academic datasets} \label{sec:academic_corpora}

Some prior datasets address CLIR in technical domains.
For example, \citet{xu2016domain} study cross-language technical question retrieval, CLEF eHealth simulates medical search by non-experts \citep{galuscakova2022clir}, and MATERIAL covers law, security, and health topics \citep{zavorin2020corpora}.
A close reference to our work is NeuCLIR 2023's technical track, which contains 40 English queries to retrieve Chinese academic abstracts across Chemistry, Economics, Physics, Biology, and Medicine \citep{lawrie2024overview}.
NeuCLIR 2024 also featured a technical task but their proceedings were unavailable at the time of writing.

Beyond CLIR-specific datasets, some parallel academic corpora similar to the one we use include academic metadata aligned across languages.
SciPar \citep{roussis2022scipar} compiles bilingual titles and abstracts from theses and dissertations.
Other examples mentioned in \citet{roussis2022scipar} include
SciELO (\citealp{neves2016scielo}, English, Portuguese, Spanish),
ASPEC (\citealp{nakazawa2016aspec}, English, Japanese, Chinese),
CAPES (\citealp{soares2018parallel}, Brazilian academic works), 
and EDP (\citealp{neveol2018parallel}, English-French biomedical texts).
In the biomedical domain, MEDLINE \citep{wu2011statistical} and BVS \citep{soares2019bvs} provide multilingual aligned abstracts.
\citet{niu2024does} introduce a dataset of translated abstracts from journals in translation studies.

These corpora mainly support MT by providing parallel abstracts and titles, often with aligned sentences.
Our work differs by using keywords as queries of a CLIR dataset. 
Among existing corpora, only CAPES and BVS include multilingual keywords suitable for this task, but they are not publicly available at the time of writing.

\section{Evaluation data} \label{sec:dataset}

\begin{figure}[ht]
  \centering
  \includegraphics[width=\linewidth]{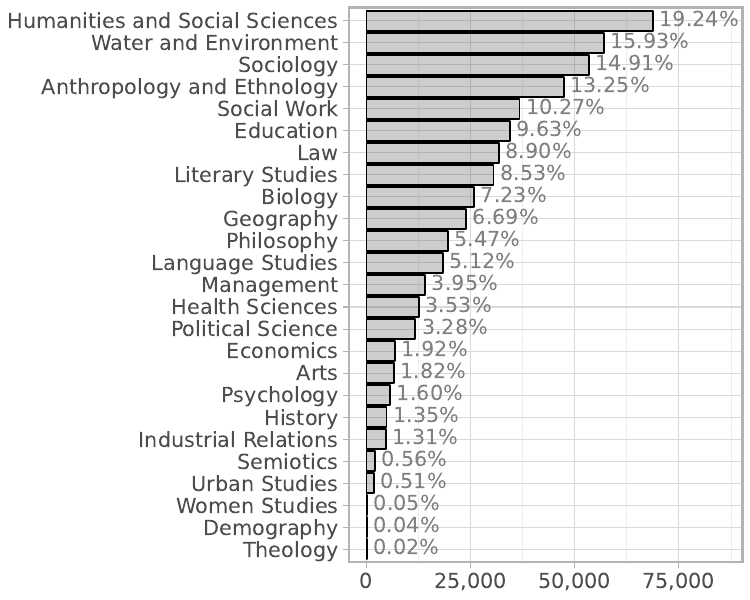}
  \caption{
    Number of queries per disciplines in the CLIRudit dataset.
    A query inherits the disciplines of the articles containing its keywords.
    Since queries can originate from multiple articles and articles can have multiple disciplines, percentages do not sum to 100\%.
  }
  \label{fig:discipline_distribution}
\end{figure}

To evaluate academic CLIR methods, we built CLIRudit using data from Érudit\footnote{\href{https://www.erudit.org/en/}{https://www.erudit.org/en/}}, a Quebec-based Canadian platform that publishes research in the arts, humanities, and social sciences.
Érudit's journals are selected by a scientific committee and meet national quality standards, ensuring the relevance and quality of the content.

We focused exclusively on research articles that included both English and French abstracts and keywords, provided by the authors.
From each article's metadata, we extracted the title, subtitle, abstract, and keywords.

Following the standard CLIR task setup, with English queries targeting non-English documents \citep{lawrie2023overview,lawrie2024overview}, we built the dataset as follows (see Fig. \ref{fig:main_figure} for an overview): 

\begin{itemize}[wide, itemindent=\labelsep] 

  \item \textbf{Queries.}  
  Created by combining all possible groups of three English keywords from each article; e.g., an article with keywords \{$A$, $B$, $C$, $D$\} generates the queries: ``$A, B, C$'', ``$A, B, D$'', ``$A, C, D$'', and ``$B, C, D$''.

  \item \textbf{Relevance judgments.}
  A document was marked relevant to a query if its English keyword metadata included all three query keywords.
  This is based on the assumption that authors try to make their work discoverable via those terms.

  \item \textbf{Document collection.}
  Each document or retrieval unit was built as the concatenation of its French title, subtitle, and abstract. 

\end{itemize}

We chose three-keyword combinations for queries based on preliminary observations.
Using only two keywords produced overly broad queries which could apply to many documents even if those specific terms weren't used by the authors; e.g., ``\textit{family dynamics, gender identity}'' or ``\textit{canada, québec}''.
In contrast, using more than three keywords led to overly narrow queries that were unlikely to reflect realistic user search behavior.

The final dataset contains 357,710 queries derived from 41,594 unique English keywords, with an average query length of 4.8 words (SD = 1.7); and 16,389 French documents from 124 journals across 25 disciplines, with an average document length of 176.7 words (SD = 82.4).
Because of the way the dataset was built, all documents in the collection are relevant to at least one query. 
99.3\% of queries have only one relevant document, showing that most three-keyword combinations are unique to a single article, which highlights the specificity of the queries.

84.9\% of the abstracts in the dataset come from articles whose primary language is French, 14.3\% from English, and 0.9\% from other languages.
The most frequent disciplines in the queries are Humanities and Social Sciences, Water and Environment, Sociology, and Anthropology and Ethnology (full distribution in Fig. \ref{fig:discipline_distribution}).

CLIRudit simulates a scenario where users know only the relevant terms in English, while the pertinent documents are only in French, with no direct translations available.
Our pipeline offers a reproducible method to build CLIR datasets for academic search.
Rather than relying on complex heuristics, it leverages the inherent bilingual structure of scientific publications.
While this work focuses on English-French retrieval, the method can be extended to other databases and language pairs, facilitating research in cross-lingual scientific retrieval.


\section{Models and methods} \label{sec:methods}

This section describes the retrieval and MT methods, and evaluation metrics used for benchmarking.

\subsection{Retrievers} \label{sec:retrievers}

We tested lexical, sparse, and dense first-stage retrievers, all operating as bi-encoders, encoding queries and documents separately.
Due to our relatively small document collection, we used exhaustive nearest-neighbor search. 
We prioritized well-documented, open-source models.

\textbf{Dense multilingual retrievers.}
We evaluated three state-of-the-art bi-encoders for direct CLIR without translation, as they are pretrained and fine-tuned on multilingual data:
\textbf{mE5}\footnote{\href{https://hf.co/intfloat/multilingual-e5-large}{intfloat/multilingual-e5-large}} \citep{wang2024multilingual}, 
\textbf{mGTE-dense}\footnote{\href{https://hf.co/Alibaba-NLP/gte-multilingual-base}{Alibaba-NLP/gte-multilingual-base}} \citep{zhang2024mgte}, and 
\textbf{BGE-m-gemma2}\footnote{\href{https://hf.co/BAAI/bge-multilingual-gemma2}{BAAI/bge-multilingual-gemma2}} \citep{li2025making}. 
While mGTE-dense and BGE-m-gemma2 are fine-tuned on some cross-lingual tasks involving mixed-language inputs, mE5 is trained on multilingual but not explicitly cross-lingual data, which may affect CLIR performance.

\textbf{Dense English retrievers.} 
We included English-focused models to assess two approaches: 
(1) retrieving French documents translated to English, or (2) leveraging cross-lingual transfer, where models, fine-tuned mainly on one language, perform well on other languages for the same task \citep{artetxe2019massively,asai2021one,zhang2023toward}.
We assessed two top English MTEB \citep{muennighoff2023mteb} performers as of early 2025:
\textbf{NV-Embed-v2}\footnote{\href{https://hf.co/nvidia/NV-Embed-v2}{nvidia/NV-Embed-v2}} \citep{lee2024nv}, and
\textbf{BGE-EN-ICL}\footnote{\href{https://hf.co/BAAI/bge-en-icl}{BAAI/bge-en-icl}} \citep{li2025making}.
Though targeting English, these models have some multilingual fine-tuning (including French), and their Mistral-7B backbone \citep{jiang2023mistral} may also have had multilingual pretraining. 
but this information is not publicly available.

\textbf{French-specialized dense retrievers.}
Few dense retrievers specialize in non-English languages, and those that do are developed by open source communities and lack thorough documentation.
We considered these top performers on the MTEB French benchmark \citep{ciancone2024mteb}:
\textbf{Croissant}\footnote{\href{https://hf.co/manu/sentence_croissant_alpha_v0.3}{manu/sentence\_croissant\_alpha\_v0.3}} (from CroissantLLM, \citealp{faysse2024croissantllm}),
\textbf{Solon}\footnote{\href{https://hf.co/OrdalieTech/Solon-embeddings-large-0.1}{OrdalieTech/Solon-embeddings-large-0.1}}, and 
\textbf{Lajavaness}\footnote{\href{https://hf.co/Lajavaness/bilingual-embedding-large}{Lajavaness/bilingual-embedding-large}},  all of which are bilingual at some degree as they include English data in pre-training or fine-tuning.

\textbf{Dense multi-vector retrievers.}
ColBERT-style models encode queries and documents into token-level embeddings, enabling fine-grained late interaction and pre-computation of document representations, with strong performance in out-of-domain retrieval \citep{khattab2020colbert, santhanam2022colbertv2}.
PLAID \citep{santhanam2022plaid} improves speed using clustering and centroid-based interaction.
We tested \textbf{PLAID-X}\footnote{\href{https://hf.co/hltcoe/plaidx-large-clef-mtd-mix-passages-mt5xxl-engeng}{plaidx-large-clef-mtd-mix-passages-mt5xxl-engeng}} \citep{yang2024distillation},
a multilingual ColBERT variant trained via translate-distill, distilling signals from an English cross-encoder and translated passages.
It uses multilingual batching to support English queries and French, German, and Spanish documents.

\textbf{Sparse retrievers.}
These encode queries and documents as term-weighted vectors, enabling efficient retrieval with inverted indexes \citep{formal2022distillation}.
We tested \textbf{BM25} \citep{robertson2009probabilistic}, a strong exact-match baseline \citep{thakur2021beir}, used on inputs translated into a common language.

Learned sparse models improve retrieval by expanding terms through supervised training \citep{lin2022simple}. 
We assessed \textbf{SPLADE++}\footnote{\href{https://hf.co/naver/splade-cocondenser-ensembledistil}{naver/splade-cocondenser-ensembledistil}} (monolingual, requires MT into English);
and the multilingual \textbf{mGTE-sparse} \citep{zhang2024mgte} and \textbf{BGE-M3-sparse} \citep{chen2024m3}, which allow cross-lingual retrieval but lack term expansion, limiting performance when queries and documents share few tokens.
We excluded BLADE \citep{nair2023blade}, a cross-lingual SPLADE variant with term expansion, due to the lack of an English-French version.
Additionally, BLADE has demonstrated lower effectiveness compared to PLAID-X, which we included in our evaluation.

Finally, we tested \textbf{PSQ} \citep{yang2024efficiency}, which enables sparse CLIR without conventional MT by indexing documents in query language tokens using a probabilistic alignment matrix \citep{yang2024translate}. 

\paragraph{}
See Appendix \ref{app:retrievers} for further details on the models and their implementations.

\subsection{Machine translation} \label{sec:machine_translation}

We tested three machine translation models:

\begin{itemize}[wide, itemindent=\labelsep, nosep]
  
  \item \textbf{GPT-4o-mini}\footnote{\href{https://openai.com/index/gpt-4o-mini-advancing-cost-efficient-intelligence/}{gpt-4o-mini}}.
  Recent work shows LLMs perform well on document-level MT \citep{kocmi2023findings, zhang2023machine,pang2025salute}.
  We used a cost-efficient proprietary model which performed competitively on high-resource language pairs \citep{hendy2023good, zhu2024multilingual}. 
  
  \item \textbf{Llama-3.2}.
  We used the 3.2B-parameter version as an open-source LLM alternative to GPT, with strong zero-shot capabilities in French to English translation \citep{zhang2023machine}.
  Open-source models can be advantageous for cost-efficiency and for the ability to fine-tune on domain-specific data.
  
  \item \textbf{OpusMT},
  a 75M-parameter French-English MarianMT encoder-decoder model \citep{tiedemann2023democratizing} trained on Opus parallel data\footnote{\href{https://hf.co/Helsinki-NLP/opus-mt-fr-en}{Helsinki-NLP/opus-mt-fr-en}}.
  While designed for sentence-level MT, we applied it at the document level following \citet{cui2024efficiently}.
  It supports up to 512 tokens, far fewer than the 100k+ limits of GPT and Llama.

\end{itemize}

For LLM translation we used a zero-shot prompt suited for instruction-tuned LLMs (details in Appendix \ref{app:translation}).
We did not test other strong proprietary translators due to lack of cost-efficient APIs.

Finally, as \textbf{gold standard} translations, we used the English translations of the French titles, subtitles, and abstracts provided by the article authors. 
These reflect the potential performance of each retrieval method using human translations.
We did not use the actual French keywords as ``gold standard'' queries since they do not map one-to-one to the English keywords; using them would alter the original set of evaluation queries and introduce noise into the analysis.

\subsection{Evaluation metrics} \label{sec:evaluation_metrics}

To measure retrieval performance, we use Recall@100 and Mean Average Precision with a 1000 cutoff rank (MAP), which have been widely used \citep{nair2023blade,lawrie2024overview,yang2024efficiency}. Whereas Recall@100 is useful to assess the effectiveness of methods when used as first-stage retrievers, MAP is more appropriate for measuring overall performance of a method used as a single-stage system \citep{yang2024efficiency}.
We compute 95\% bootstrap confidence intervals with 1,000 resamples to assess statistical significance.

To evaluate document translation quality, we used three metrics used in recent works \citep{sun2022rethinking,zhang2022multilingual,zhuocheng2023addressing}: BLONDE \citep{jiang2022blonde}, document-BLEU (d-BLEU, \citealp{liu2020multilingual}), and document-chrF (d-chrF, \citealp{zhuocheng2023addressing}).

\section{Results and analysis} \label{sec:results}

\begin{table*}[ht!]
  \small
  \centering
\begin{tabular}{
    l@{\hspace{6pt}} | 
    @{\hspace{6pt}}c @{\hspace{6pt}}c @{\hspace{6pt}}c @{\hspace{6pt}}c
    @{\hspace{6pt}}c@{\hspace{6pt}}c | 
    c@{\hspace{6pt}}c@{\hspace{6pt}}c@{\hspace{6pt}}c@{\hspace{6pt}}c@{\hspace{6pt}}c
    }
    \toprule
    & \multicolumn{6}{c|}{\textbf{MAP}} & \multicolumn{6}{c}{\textbf{Recall@100}} \\
    \cmidrule(lr{6pt}){2-7} \cmidrule(lr{6pt}){8-13}
    \multirow{2}{*}{\textbf{\begin{tabular}[c]{@{}l@{}}Machine\\ Trans. ($\rightarrow$)\end{tabular}}} & 
    \multirow{3}{*}{\textbf{None}} & 
    \multirow{3}{*}{\textbf{\begin{tabular}[c]{@{}c@{}}Query\\ (GPT4)\end{tabular}}} & 
    \multicolumn{3}{c}{\multirow{2}{*}{\textbf{Doc.}}} & 
    \multirow{3}{*}{\textbf{\begin{tabular}[c]{@{}c@{}}Gold\end{tabular}}} &
    \multirow{3}{*}{\textbf{None}} & 
    \multirow{3}{*}{\textbf{\begin{tabular}[c]{@{}c@{}}Query\\ (GPT4)\end{tabular}}} & 
    \multicolumn{3}{c}{\multirow{2}{*}{\textbf{Doc.}}} & 
    \multirow{3}{*}{\textbf{\begin{tabular}[c]{@{}c@{}}Gold\end{tabular}}} \\
     &  &  & \multicolumn{3}{c}{} &  &  &  & \multicolumn{3}{c}{} &  \\ \cmidrule(lr{6pt}){4-6} \cmidrule(lr{6pt}){10-12}
    \textbf{Retriever ($\downarrow$)} &  &  & 
    \textbf{Opus} & 
    \textbf{Llama} & 
    \textbf{GPT4} & 
    &  &  & 
    \textbf{Opus} & 
    \textbf{Llama} & 
    \textbf{GPT4} &  \\
    \midrule

    mE5 & 
    0.434 & 0.412 & 0.448 & 0.480 & \underline{0.490} & 0.526 & 
    0.784 & 0.760 & 0.790 & 0.817 & \underline{0.823} & 0.840 \\
    mGTE-dense & 
    0.450 & 0.445 & 0.452 & 0.459 & \underline{0.468} & 0.496 & 
    0.820 & 0.813 & 0.820 & 0.834 & \underline{0.837} & 0.849 \\
    BGE-m-gemma2 & 
    \underline{0.571} & 0.543 & 0.533 & 0.548 & 0.560 & 0.571 & 
    \textbf{0.903} & \textbf{0.895} & \textbf{0.894} & \textbf{0.908} & \textbf{\underline{0.910}} & \textbf{0.917} \\
    NV-Embed-v2 & 
    \textbf{0.580} & \textbf{0.575} & \textbf{0.541} & 0.569 & \textbf{\underline{0.586}} & 0.600 & 
    \underline{0.895} & 0.889 & 0.866 & 0.887 & 0.892 & 0.894 \\
    BGE-EN-ICL & 
    \underline{0.507} & 0.441 & 0.411 & 0.486 & 0.501 & 0.535 & 
    \underline{0.857} & 0.810 & 0.760 & 0.831 & 0.837 & 0.861 \\
    Croissant & 
    0.358 & \underline{0.365} & 0.325 & 0.345 & 0.357 & 0.376 & 
    0.793 & \underline{0.794} & 0.748 & 0.773 & 0.781 & 0.794 \\
    Solon & 
    0.507 & 0.516 & 0.502 & 0.520 & \underline{0.536} & 0.555 & 
    0.856 & 0.858 & 0.845 & 0.860 & \underline{0.866} & 0.870 \\
    Lajavaness & 
    \underline{0.472} & 0.454 & 0.431 & 0.457 & 0.470 & 0.486 & 
    \underline{0.848} & 0.838 & 0.817 & 0.836 & 0.843 & 0.849 \\
    PLAID-X & 
    0.578 & 0.548 & 0.539 & \textbf{0.572} & \textbf{\underline{0.586}} & 0.605 & 
    0.870 & 0.854 & 0.845 & 0.869 & \underline{0.874} & 0.879 \\
    SPLADE++ & 
    0.284 & 0.426 & 0.530 & 0.548 & \underline{0.572} & 0.609 & 
    0.604 & 0.753 & 0.836 & 0.853 & \underline{0.864} & 0.875 \\
    mGTE-sparse & 
    0.169 & \underline{0.434} & 0.401 & 0.405 & 0.428 & 0.487 & 
    0.443 & 0.763 & 0.737 & 0.760 & \underline{0.771} & 0.805 \\
    BGE-M3-sparse & 
    0.177 & 0.458 & 0.413 & 0.434 & \underline{0.460} & 0.511 & 
    0.449 & \underline{0.781} & 0.738 & 0.763 & 0.778 & 0.807 \\
    BM25 & 
    0.181 & 0.390 & 0.488 & 0.513 & \underline{0.549} & \textbf{0.611} & 
    0.417 & 0.706 & 0.789 & 0.815 & \underline{0.832} & 0.861 \\
    PSQ & 
    \underline{0.440} & - & - & - & - & - & 
    \underline{0.756} & - & - & - & - & - \\
    
    \bottomrule
\end{tabular}

  \caption{
    MAP and Recall@100 in CLIRudit.
    Best column scores are in bold; best row scores per metric are underlined, excluding gold translation.
    Statistical significance is shown in Appendix \ref{app:significance} for better readability.
  }
  \label{tab:metrics}
\end{table*}


This section analyzes the performance of retrieval and translation models on CLIRudit (Table \ref{tab:metrics}). 
Due to the large sample size, no confidence interval width exceeded 0.003. 
Intervals are omitted here for readability (see Appendix \ref{app:significance}).
To account for input length effects, we evaluated each method both at its native input limit and with the same 512-token limit. 
Results differed by no more than 0.005 from the reported values, small enough to not affect general trends.

We now discuss key findings from the results.

\textbf{1. Without translation, dense retrievers excel, even without multilingual retrieval fine-tuning.}
NV-Embed-v2 and PLAID-X achieved the highest MAP, while BGE-m-gemma2 led in Recall@100.
Interestingly, NV-Embed-v2 is not reported to have multilingual capabilities; though its fine-tuning data, which is English-only for retrieval, includes French in STS17 and STS22 sentence pairs \citep{cer2017semeval, chen2022semeval}.

BM25 with the French analyzer performed poorly without MT due to the query-document language mismatch, but still had non-zero results.
This shows CLIRudit has some query-document lexical overlap; manual inspection revealed shared terms like proper nouns, Latin terms, and acronyms.

Among sparse models, SPLADE++ outperformed mGTE-sparse and BGE-M3-sparse, likely thanks to query expansion mitigating the language mismatch. 
PSQ addresses this mismatch via probabilistic translation, reaching MAP comparable to larger dense models like mE5 and mGTE-dense.

\begin{figure*}[ht]
  \centering
  \includegraphics[width=\linewidth]{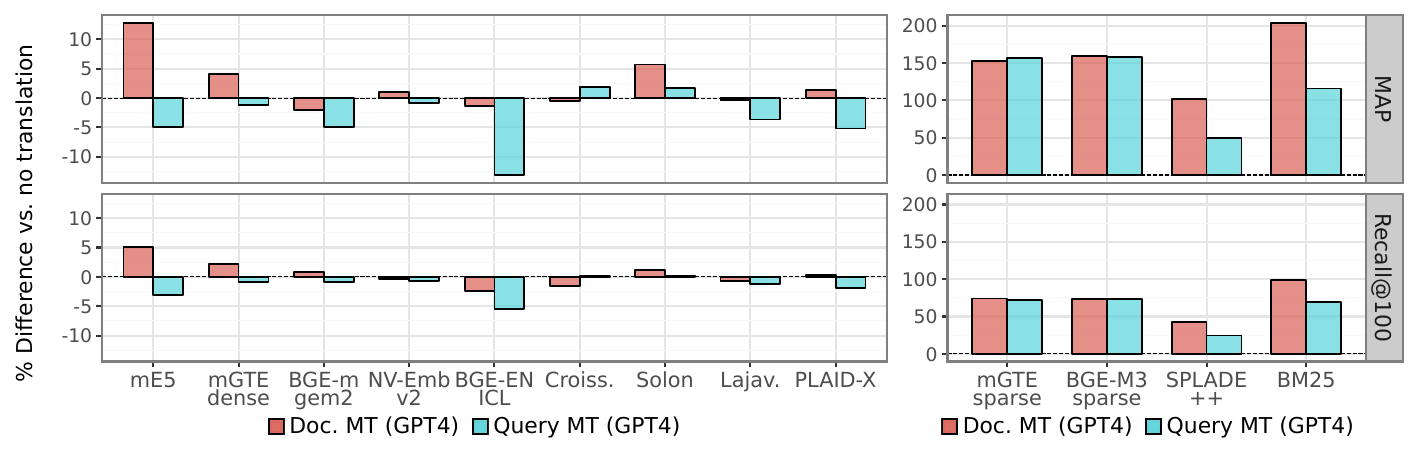}
  \caption{
    \% difference in MAP and Recall@100 of document translation (red) and query translation (blue) compared to no translation.
    Positive (negative) values indicate improvement (degradation) with translation.
    For ease of visualization, sparse models are shown with a different scale and only GPT translation is considered.
  }
  \label{fig:translation_diff}
\end{figure*}

\textbf{2. Document translation can improve dense retrievers.}
DT with GPT-4o-mini improved dense retriever MAP by up to 10\% and Recall@100 by up to 5\% (Fig. \ref{fig:translation_diff}, left). 
The highest MAP overall came from NV-Embed-v2+DT and PLAID-X+DT with GPT-4o-mini.
However, translation sometimes hurt performance, especially with QT, affecting models like mE5, BGE-EN-ICL, and even top-performing ones like NV-Embed-v2 and PLAID-X. 

Manual review showed QT can reduce recall by mistranslating proper nouns with identical cross-language spelling. 
For example, ``\textit{Goose Bay}'' (a Canadian town) was incorrectly translated as ``\textit{Baie aux Oies}'' instead of remaining unchanged.


\textbf{3. Document translation usually outperformed query translation for sparse retrievers.}
Translation had a modest effect on dense models but significantly boosted sparse retrieval. 
Moreover, DT consistently outperformed QT (Fig. \ref{fig:translation_diff} right), especially for BM25 and SPLADE++, with SPLADE++ plus DT nearing the top dense retriever MAP, and also outperforming the PSQ probabilistic translation method (Table \ref{tab:metrics}).

While DT may offer richer context than QT \citep{galuscakova2022clir,lin2022simple}, DT outperforming QT is expected for SPLADE++ since it's trained only in English.
In contrast, mGTE-sparse and BGE-M3-sparse performed similarly with QT and DT.

Manual inspection of BM25 cases where DT outperformed QT shows that DT can preserve key terms better.
For example, ``\textit{fair innings}'' correctly remains unchanged with DT to English, but translating the query to French yields ``\textit{juste part}'', which isn't in the original document.
Similarly, the term ``\textit{beck}'' in a query about the surname of a social scientist is correctly preserved in DT, but mistranslated as ``\textit{appel}'' in the query (French for ``call''), making the document irretrievable.

\begin{figure}[ht!]
  \centering
  \includegraphics[width=\linewidth]{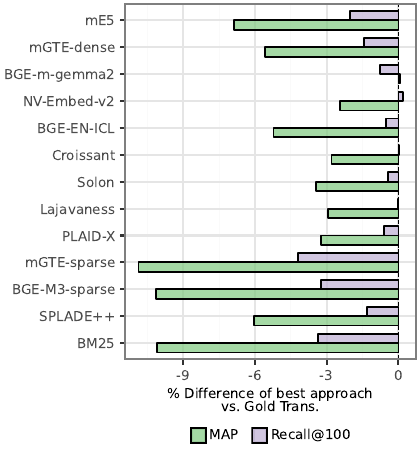}
  \caption{
    \% difference in MAP (green) and Recall@100 (purple) for the best-performing approach of each retriever, relative to gold-standard translations.
  }
  \label{fig:gold_diff}
\end{figure}

\textbf{4. Document translation quality correlated with retrieval performance.}
GPT-4o-mini led in document translation quality (BLEU=34.41, BLONDE=49.32, chrF=63.83), followed closely by Llama (BLEU=31.27, BLONDE=46.52, chrF=61.56), with OpusMT trailing far behind (BLEU: 10.77, BLONDE: 19.35, chrF: 36.15). 
This ranking mirrors their retrieval performance, where GPT-4o-mini systematically outperformed Llama, which in turn outperformed OpusMT (Table \ref{tab:metrics}).
While these results indicate a correlation between translation and retrieval quality, quantifying MT's exact contribution requires further study beyond the scope of this paper.

\textbf{5. Top dense retrievers approached gold translation recall.} 
Models like NV-Embed-v2, BGE-m-gemma2, BGE-EN-ICL, and PLAID-X, performed close to their gold translation recall (Fig. \ref{fig:gold_diff}). 
Except for BGE-m-gemma2, gaps in MAP were larger, indicating potential for better ranking.

Sparse models BM25 and SPLADE++ achieved the highest MAP with gold translations (Table \ref{tab:metrics}), highlighting the impact of translation quality. 
Because CLIRudit queries are keywords and documents are abstracts, sparse models naturally perform well with accurate translations.
SPLADE's smaller gap to gold as compared to other sparse methods suggests greater robustness to translation errors, likely due to query expansion.

\begin{figure}[ht!]
  \centering
  \includegraphics[width=\linewidth]{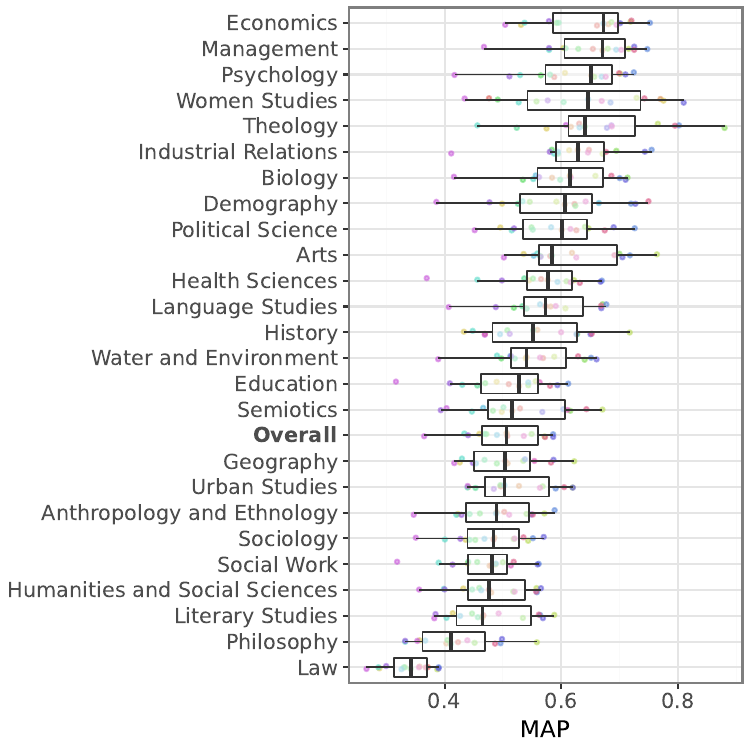}
  \caption{
    MAP of retrievers across CLIRudit disciplines. 
    Each dot represents a method's MAP in a discipline's queries, using its best translation method (excluding gold).
    Dot colors indicate retrievers: Croissant (pink) often performs worst, while the best varies by discipline.
    }
    \label{fig:disciplines_performance}
\end{figure}
  
\textbf{6. Performance varies significantly across disciplines.}
Considering the best-performing approach for each retriever, MAP was on average higher in Industrial Relations, Theology, Women's Studies, Psychology, Management, and Economics, and lower in Philosophy and Law (Fig. \ref{fig:disciplines_performance}). 
While Croissant was typically the weakest across disciplines, no translation-retriever combination consistently outperformed the others.

\section{Discussion} \label{sec:discussion}

Dense single-vector retrievers based on large decoder-only models (e.g., NV-Embed-v2, BGE-m-gemma2) achieve near gold translation-augmented performance without additional training, which may result from pretraining on large corpora and cross-lingual transfer capabilities.
A smaller, CLIR-specialized model, PLAID-X, also performed competitively; at the expense of needing language- and task-specific training data and having higher search latency due to its multi-vector design \citep{santhanam2022plaid}.
Both dense approaches avoid the overhead of translating the entire corpus, but large models may incur high indexing costs on large collections.

Sparse retrievers, lexical or learned, offer faster indexing and search, but need translation to narrow the gap with dense methods, and still fall short in overall performance. 
DT outperformed QT, likely because it provides richer context; and it can be done offline, which is important when using costly MT systems.
QT enables quicker experimentation by avoiding corpus reindexing with each new model, but usually with lower accuracy \citep{lin2022simple, galuscakova2022clir}. 
Ultimately, the choice of method comes down to balancing retrieval performance, indexing and search latency, and translation costs.

Our dataset uses keyword-based queries, reflecting how authors describe their work to make it discoverable. 
This assumes users know the right keywords, shifting the dataset challenge to language differences in a technical domain rather than query formulation.
This allows meaningful analysis, though it's unclear how system rankings might change with other types of queries, e.g., natural language questions.
Our approach aligns with other datasets using non-natural or generated queries, such as SCIDOCS, DBPedia \citep{thakur2021beir}, WikiCLIR \citep{sasaki2018cross}, and CLIRMatrix \citep{sun2020clirmatrix}.

Like all IR datasets, ours has limitations in scope and collection method, so we encourage evaluation on many, diverse datasets.
As the first English-French academic retrieval dataset, CLIRudit adds to this diversity and complements existing resources.

\section{Conclusions} \label{sec:conclusions}

We introduced a method for building CLIR datasets from bilingual metadata in scientific publications.
By using keywords as queries and abstracts as documents, this approach enables automated, scalable creation of large evaluation resources without manual annotation or complex heuristics. 
We applied it to produce CLIRudit, the first English-French CLIR dataset for academic search, based on a real-world database.

Evaluations of single-stage methods on CLIRudit showed that: (1) state-of-the-art dense bi-encoders achieved strong cross-lingual performance without translation, nearing monolingual retrieval with gold translations; (2) sparse retrievers with document translation were competitive; and (3) document translation generally outperformed query translation, likely due to richer context.

These results have practical implications for academic search systems.
Large dense retrievers deliver the best performance, but the strong results of sparse retrievers with document translation suggest a viable alternative that may be more practical to implement at scale.
This is particularly relevant for academic publishing platforms like Érudit that aim to make their content more discoverable to researchers.

Our method can be applied to other academic databases and language pairs, supporting broader research in cross-lingual access to scientific knowledge.

\section*{Limitations}

Our dataset's document collection includes only relevant documents, unlike in real applications where relevant documents might coexist with a much larger collection. 
The values reported may not be representative of real-world settings. 
The reported metrics should be used to compare methods rather than to provide absolute performance estimates, which is standard practice in IR research \citep{thakur2021beir}.

Our dataset may also contain some false negatives: some relevant documents may not be labeled as such if some authors did not include some suitable keywords in the metadata, while others did.
However, because queries consist of three keywords, they are relatively specific, likely reducing false negatives, as it is unlikely that there is more than one document in the collection relevant to a narrow query. 

We found that the proprietary GPT-4o-mini LLM outperformed the open-source Llama 3.2 and the smaller OpusMT encoder-decoder for zero-shot translation.
Further exploration with few-shot prompting or fine-tuning may improve the performance of the open-source models.
In addition, OpusMT is not optimized for document translation, so using sentence-level translation may be more optimal.
However, this approach requires a more complex pipeline with sentence splitting and risks losing cross-sentence coherence.

Possible data contamination is a concern for fair evaluation: our test set may appear in the training data of pre-trained models, especially LLMs used for translation and retrievers initialized from LLMs, such as NV-Embed-v2 and BGE-m-gemma2.
This could lead to inflated results, but is difficult to verify due to the lack of information about the exact training data of these models \citep{sainz2023nlp,oren2024proving}.

Our dataset is limited to keyword-based queries and metadata-only documents. 
Results may differ with other query types, e.g. natural language questions, or full-text documents. Future work could explore approaches that use other types of queries or full-text representations.
We also focused on French, a high-resource language; performance may vary in low-resource settings due to lower translation quality and limited training data for retrievers.

We tested single-stage retrieval without re-ranking, fusion, or pseudo-relevance feedback \citep{lin2022simple}. Including these techniques could enhance performance and reveal additional insights into CLIR system design. 
We also did not analyze the computational costs of translation, retrieval, or indexing, as explored in prior work \citep{rosa2021cost,nair2023blade}. Such analysis would be valuable for assessing the trade-offs between effectiveness and efficiency in practical deployment.
Additionally, we did not fine-tune or train any retrieval models on our dataset. Training on domain-specific data could potentially lead to better performance, both on our dataset and on others.

\section*{Acknowledgments}

This project was funded by the Social Science and Humanities Research Council of Canada Pan-Canadian Knowledge Access Initiative Grant (Grant 1007-2023-0001), and the Fonds de recherche du Québec-Société et Culture through the Programme d'appui aux Chaires UNESCO (Grant 338828).

We used computational resources from NodoIA San Francisco (Ministry of Science and Technology of the Province of Córdoba, Argentina).

We thank Érudit for their support and access to data.

\bibliography{custom}

\appendix

\section{Retrievers} \label{app:retrievers}

\begin{table*}[ht!]
  \small
  \centering
  \begin{tabular}{ l| L | l l| c c c }
    \toprule
    \textbf{Retriever} & \textbf{Type} & \textbf{Pre-train. Lang.} & \textbf{Fine-tuning Lang.} & \textbf{\#Params.} & \textbf{Emb. Dim.} & \textbf{Max. Len.} \\ 
    \midrule
 
    mE5 & \multirow{8}{*}{\makecell{Dense\\(single-\\vector)}} & \multirow{5}{*}{Multilingual} & Mostly English & 560M & 1024 & 512 \\
    mGTE-dense &  &   & Mostly English, Chinese & 305M & 768 & 8192 \\
    BGE-m-gemma2 &  &    & Mostly English, Chinese & 9.2B & 3584 & 8192 \\
    Solon &  &  & French                             & 560M & 1024 & 512 \\
    Lajavaness &  &  & French-English                        & 560M & 1024 & 512 \\ 
    \cmidrule(lr){3-4}
    
    Croissant &  & French-English & French-English    & 1.3B & 2048 & 2048 \\
    \cmidrule(lr){3-4}
    
    NV-Embed-v2  &    & \multirow{2}{*}{Unknown} & Mostly English & 7.8B & 4096 & 32768 \\
    BGE-EN-ICL  &  &                          & Mostly English & 7.1B & 4096 & 512 \\
    \midrule

    PLAID-X  & Dense (multi-vector) & Multilingual & \multirow{1}{*}{\makecell{English, French,\\German, Spanish}} & 560M & \multirow{1}{*}{\makecell{128\\per token}} & 512 \\     
    \midrule

    mGTE-sparse  & \multirow{3}{*}{\makecell{Sparse\\(Learned)}}   & \multirow{2}{*}{\makecell{Multilingual}}  & Mostly English, Chinese  & 305M & 250,000\textsuperscript{*} & 8192 \\
    BGE-M3-sparse  &   &   & Mostly English, Chinese & 568M & 250,000\textsuperscript{*} & 8192 \\ 
    \cmidrule(lr){3-4}

    SPLADE++ &  & English & English          & 110M & 30,522\textsuperscript{*} & 512 \\
    \midrule

    BM25  & \multirow{2}{*}{\makecell{Sparse\\(Lexical)}} & -- & -- & -- & 49,144\textsuperscript{*} & -- \\
    PSQ  &                                               & -- & -- & -- & 715,837\textsuperscript{*} & -- \\ 
 
    \bottomrule
\end{tabular}

  \caption{
    Retrievers used in the study.
    \#Params.: Number of parameters.
    Emb. Dim.: Document embedding dimension.
    Max. Len.: Maximum number of input tokens allowed by the model.
    The values in the pretraining and fine-tuning language columns mentioned are approximations; in many cases, intermediate steps are involved, such as initializing from a pretrained model, followed by training with weak supervision and supervised fine-tuning. 
    However, in all cases, fine-tuning data includes some degree of French data. 
    The specific checkpoints used are given in footnotes in section \ref{sec:retrievers}.
    \newline\textsuperscript{*}The embedding dimension of sparse methods is the underlying vocabulary size.
    }
    \label{tab:retrievers_overview}
\end{table*}

Table \ref{tab:retrievers_overview} provides an overview of the retrievers evaluated in our study.

Inference with neural models was run using 16-bit floating point (fp16) inference on two NVIDIA A30 GPUs, each with 24GB of memory.

BGE-EN-ICL, BGE-m-gemma2, and NV-Embed-v2 require appending task-specific instructions before encoding the queries, which we did following the authors' templates. BGE-EN-ICL \citep{li2025making} was used in its zero-shot mode, i.e., without in-context examples appended to the queries.

We also experimented with BGE-M3-dense\footnote{\href{https://hf.co/BAAI/bge-m3}{https://hf.co/BAAI/bge-m3}} \citep{chen2024m3}, which we excluded from the body of the paper because it did not show improved performance or valuable insights.

We implemented BM25 using Pyserini with default parameters and language-specific analyzers \citep{lin2021pyserini}. For PSQ, we used the \texttt{fast\_psq} implementation by \citet{yang2024efficiency}\footnote{\href{https://github.com/hltcoe/PSQ}{https://github.com/hltcoe/PSQ}} with default parameters. We used the English-French matrix trained on 17.6M parallel sentences provided by \citet{yang2024efficiency}.

\section{Translation} \label{app:translation}

For LLM-based translation, we used a zero-shot prompt inspired by established best practices for instruction-tuned LLMs\footnote{
  \href{https://docs.anthropic.com/en/prompt-library/polyglot-superpowers}{https://docs.anthropic.com/en/prompt-library/polyglot-superpowers},
  {\href{https://platform.openai.com/docs/examples/default-translation}{https://platform.openai.com/docs/examples/default-translation}}.
}.
The complete prompt is provided in Table \ref{tab:translation_prompt}. We used sampling with 0.1 temperature and 1.0 top-p.

\begin{table}[ht]
  \small
  \begin{flushleft}
    \texttt{You are a highly skilled translator from French to English.
\\Your task is to accurately translate the French text I provide into English. 
\\You will be provided with a text, and you will output a JSON object containing the following information:
\\
\{
\\
\, \, translation: string // the translated text
\\
\}
\\
Preserve the meaning, tone, and nuance of the original text.
\\Please maintain proper grammar, spelling, and punctuation in the translated version.}

  \end{flushleft}
  \caption{Prompt used for document translation with LLMs. We used a slight variation of this prompt for query translation.}
  \label{tab:translation_prompt}
\end{table}

\section{Statistical signficance} \label{app:significance}

Tables \ref{tab:ci_map} and \ref{tab:ci_recall} show the 95\% bootstrap confidence intervals for MAP and Recall@100, respectively, for each retrieval method and translation method. 
For better readability, instead of showing the lower and upper bounds of the confidence intervals, we choose to show which systems are non-significantly different from each other, i.e., the intervals overlap. 
For example, the MAP interval of PLAID-X+DT with GPT4 overlaps with the interval of NV-Embed-v2+DT with GPT4, but not with the interval of NV-Embed-v2 with no translation (Table \ref{tab:ci_map}).

\begin{table*}[ht]
  \small
  \centering
  \begin{adjustbox}{valign=t}
  \begin{minipage}{0.49\textwidth}
  \centering
  \begin{tabular}{llll}
\toprule
Retriever & Translation &  & MAP \\
\midrule
BM25 & Gold & 1 & 0.611\textsuperscript{ 2} \\
SPLADE++ & Gold & 2 & 0.609\textsuperscript{ 1} \\
PLAID-X & Gold & 3 & 0.605 \\
NV-Embed-v2 & Gold & 4 & 0.600 \\
PLAID-X & Docs. (G4) & 5 & 0.586\textsuperscript{ 6} \\
NV-Embed-v2 & Docs. (G4) & 6 & 0.586\textsuperscript{ 5} \\
NV-Embed-v2 & None & 7 & 0.580\textsuperscript{ 8} \\
PLAID-X & None & 8 & 0.578\textsuperscript{ 7,9} \\
NV-Embed-v2 & Query (G4) & 9 & 0.575\textsuperscript{ 8} \\
SPLADE++ & Docs. (G4) & 10 & 0.572\textsuperscript{ 11,12,13} \\
PLAID-X & Docs. (L3) & 11 & 0.572\textsuperscript{ 10,12,13,14} \\
BGE-m-gemma2 & None & 12 & 0.571\textsuperscript{ 10,11,13,14} \\
BGE-m-gemma2 & Gold & 13 & 0.571\textsuperscript{ 10,11,12,14} \\
NV-Embed-v2 & Docs. (L3) & 14 & 0.569\textsuperscript{ 11,12,13} \\
BGE-m-gemma2 & Docs. (G4) & 15 & 0.560 \\
Solon & Gold & 16 & 0.555 \\
BM25 & Docs. (G4) & 17 & 0.549\textsuperscript{ 18,19,20} \\
SPLADE++ & Docs. (L3) & 18 & 0.548\textsuperscript{ 17,19,20} \\
PLAID-X & Query (G4) & 19 & 0.548\textsuperscript{ 17,18,20} \\
BGE-m-gemma2 & Docs. (L3) & 20 & 0.548\textsuperscript{ 17,18,19} \\
BGE-m-gemma2 & Query (G4) & 21 & 0.543\textsuperscript{ 22} \\
NV-Embed-v2 & Docs. (Op) & 22 & 0.541\textsuperscript{ 21,23} \\
PLAID-X & Docs. (Op) & 23 & 0.539\textsuperscript{ 22} \\
Solon & Docs. (G4) & 24 & 0.536\textsuperscript{ 25} \\
BGE-EN-ICL & Gold & 25 & 0.535\textsuperscript{ 24,26} \\
BGE-m-gemma2 & Docs. (Op) & 26 & 0.533\textsuperscript{ 25,27} \\
SPLADE++ & Docs. (Op) & 27 & 0.530\textsuperscript{ 26} \\
mE5 & Gold & 28 & 0.526 \\
Solon & Docs. (L3) & 29 & 0.520 \\
Solon & Query (G4) & 30 & 0.516\textsuperscript{ 31} \\
BM25 & Docs. (L3) & 31 & 0.513\textsuperscript{ 30,32} \\
BGE-M3-sparse & Gold & 32 & 0.511\textsuperscript{ 31} \\
BGE-EN-ICL & None & 33 & 0.507\textsuperscript{ 34} \\
Solon & None & 34 & 0.507\textsuperscript{ 33} \\
Solon & Docs. (Op) & 35 & 0.502\textsuperscript{ 36} \\
BGE-EN-ICL & Docs. (G4) & 36 & 0.501\textsuperscript{ 35} \\
mGTE-dense & Gold & 37 & 0.496 \\
mE5 & Docs. (G4) & 38 & 0.490\textsuperscript{ 39,40} \\
BM25 & Docs. (Op) & 39 & 0.488\textsuperscript{ 38,40,41,42} \\
\bottomrule
\end{tabular}

    \end{minipage}
    \end{adjustbox}
    \hfill
    \begin{adjustbox}{valign=t}
    \begin{minipage}{0.49\textwidth}
    \centering
    
\begin{tabular}{llll}
\toprule
Retriever & Translation &  & MAP \\
\midrule
mGTE-sparse & Gold & 40 & 0.487\textsuperscript{ 38,39,41,42} \\
BGE-EN-ICL & Docs. (L3) & 41 & 0.486\textsuperscript{ 39,40,42} \\
Lajavaness & Gold & 42 & 0.486\textsuperscript{ 39,40,41} \\
mE5 & Docs. (L3) & 43 & 0.480 \\
Lajavaness & None & 44 & 0.472\textsuperscript{ 45} \\
Lajavaness & Docs. (G4) & 45 & 0.470\textsuperscript{ 44,46} \\
mGTE-dense & Docs. (G4) & 46 & 0.468\textsuperscript{ 45} \\
BGE-M3-sparse & Docs. (G4) & 47 & 0.460\textsuperscript{ 48,49,50} \\
mGTE-dense & Docs. (L3) & 48 & 0.459\textsuperscript{ 47,49,50} \\
BGE-M3-sparse & Query (G4) & 49 & 0.458\textsuperscript{ 47,48,50} \\
Lajavaness & Docs. (L3) & 50 & 0.457\textsuperscript{ 47,48,49} \\
Lajavaness & Query (G4) & 51 & 0.454 \\
mGTE-dense & Docs. (Op) & 52 & 0.452\textsuperscript{ 53} \\
mGTE-dense & None & 53 & 0.450\textsuperscript{ 52,54} \\
mE5 & Docs. (Op) & 54 & 0.448\textsuperscript{ 53} \\
mGTE-dense & Query (G4) & 55 & 0.445 \\
BGE-EN-ICL & Query (G4) & 56 & 0.441 \\
BGE-M3-sparse & Docs. (L3) & 57 & 0.434\textsuperscript{ 58,59,60} \\
mGTE-sparse & Query (G4) & 58 & 0.434\textsuperscript{ 57,59,60} \\
mE5 & None & 59 & 0.434\textsuperscript{ 57,58,60} \\
Lajavaness & Docs. (Op) & 60 & 0.431\textsuperscript{ 57,58,59} \\
mGTE-sparse & Docs. (G4) & 61 & 0.428\textsuperscript{ 62} \\
SPLADE++ & Query (G4) & 62 & 0.426\textsuperscript{ 61} \\
BGE-M3-sparse & Docs. (Op) & 63 & 0.413\textsuperscript{ 64,65} \\
mE5 & Query (G4) & 64 & 0.412\textsuperscript{ 63,65} \\
BGE-EN-ICL & Docs. (Op) & 65 & 0.411\textsuperscript{ 63,64} \\
mGTE-sparse & Docs. (L3) & 66 & 0.405 \\
mGTE-sparse & Docs. (Op) & 67 & 0.401 \\
BM25 & Query (G4) & 68 & 0.390 \\
Croissant & Gold & 69 & 0.376 \\
Croissant & Query (G4) & 70 & 0.365 \\
Croissant & None & 71 & 0.358\textsuperscript{ 72} \\
Croissant & Docs. (G4) & 72 & 0.357\textsuperscript{ 71} \\
Croissant & Docs. (L3) & 73 & 0.345 \\
Croissant & Docs. (Op) & 74 & 0.325 \\
SPLADE++ & None & 75 & 0.284 \\
BM25 & None & 76 & 0.181 \\
BGE-M3-sparse & None & 77 & 0.177 \\
mGTE-sparse & None & 78 & 0.169 \\
PSQ & None & 79 & 0.123 \\
\bottomrule
\end{tabular}

  \end{minipage}
  \end{adjustbox}
  \caption{
    95\% bootstrap confidence intervals for MAP, using 1000 resamples.
    Numbers in subscripts indicate the 95\% interval of the system of the row overlaps with the interval of the systems in the subscripts.
    \\G4: GPT-4o-mini. L3: Llama-3.2. Op: OpusMT. 
  }
  \label{tab:ci_map}
\end{table*}

\begin{table*}[ht]
  \small
  \centering
  \begin{adjustbox}{valign=t}
  \begin{minipage}{0.49\textwidth}
  \centering
  \begin{tabular}{llll}
\toprule
Retriever & Translation &  & Recall@100 \\
\midrule
BGE-m-gemma2 & Gold & 1 & 0.917 \\
BGE-m-gemma2 & Docs. (G4) & 2 & 0.910 \\
BGE-m-gemma2 & Docs. (L3) & 3 & 0.908 \\
BGE-m-gemma2 & None & 4 & 0.903 \\
NV-Embed-v2 & None & 5 & 0.895\textsuperscript{ 6,7,8} \\
BGE-m-gemma2 & Query (G4) & 6 & 0.895\textsuperscript{ 5,7,8} \\
BGE-m-gemma2 & Docs. (Op) & 7 & 0.894\textsuperscript{ 5,6,8} \\
NV-Embed-v2 & Gold & 8 & 0.894\textsuperscript{ 5,6,7,9} \\
NV-Embed-v2 & Docs. (G4) & 9 & 0.892\textsuperscript{ 8} \\
NV-Embed-v2 & Query (G4) & 10 & 0.889 \\
NV-Embed-v2 & Docs. (L3) & 11 & 0.887 \\
PLAID-X & Gold & 12 & 0.879 \\
SPLADE++ & Gold & 13 & 0.875\textsuperscript{ 14} \\
PLAID-X & Docs. (G4) & 14 & 0.874\textsuperscript{ 13} \\
PLAID-X & None & 15 & 0.870\textsuperscript{ 16,17} \\
Solon & Gold & 16 & 0.870\textsuperscript{ 15,17} \\
PLAID-X & Docs. (L3) & 17 & 0.869\textsuperscript{ 15,16} \\
Solon & Docs. (G4) & 18 & 0.866\textsuperscript{ 19} \\
NV-Embed-v2 & Docs. (Op) & 19 & 0.866\textsuperscript{ 18,20} \\
SPLADE++ & Docs. (G4) & 20 & 0.864\textsuperscript{ 19} \\
BGE-EN-ICL & Gold & 21 & 0.861\textsuperscript{ 22,23} \\
BM25 & Gold & 22 & 0.861\textsuperscript{ 21,23} \\
Solon & Docs. (L3) & 23 & 0.860\textsuperscript{ 21,22,24} \\
Solon & Query (G4) & 24 & 0.858\textsuperscript{ 23,25,26} \\
BGE-EN-ICL & None & 25 & 0.857\textsuperscript{ 24,26} \\
Solon & None & 26 & 0.856\textsuperscript{ 24,25,27} \\
PLAID-X & Query (G4) & 27 & 0.854\textsuperscript{ 26,28} \\
SPLADE++ & Docs. (L3) & 28 & 0.853\textsuperscript{ 27} \\
mGTE-dense & Gold & 29 & 0.849\textsuperscript{ 30,31} \\
Lajavaness & Gold & 30 & 0.849\textsuperscript{ 29,31} \\
Lajavaness & None & 31 & 0.848\textsuperscript{ 29,30} \\
PLAID-X & Docs. (Op) & 32 & 0.845\textsuperscript{ 33} \\
Solon & Docs. (Op) & 33 & 0.845\textsuperscript{ 32,34} \\
Lajavaness & Docs. (G4) & 34 & 0.843\textsuperscript{ 33} \\
mE5 & Gold & 35 & 0.840 \\
Lajavaness & Query (G4) & 36 & 0.838\textsuperscript{ 37,38,39,40} \\
mGTE-dense & Docs. (G4) & 37 & 0.837\textsuperscript{ 36,38,39,40} \\
BGE-EN-ICL & Docs. (G4) & 38 & 0.837\textsuperscript{ 36,37,39,40} \\
Lajavaness & Docs. (L3) & 39 & 0.836\textsuperscript{ 36,37,38,40,41} \\
\bottomrule
\end{tabular}

    \end{minipage}
    \end{adjustbox}
    \hfill
    \begin{adjustbox}{valign=t}
    \begin{minipage}{0.49\textwidth}
    \centering
    
\begin{tabular}{llll}
\toprule
Retriever & Translation &  & Recall@100 \\
\midrule
SPLADE++ & Docs. (Op) & 40 & 0.836\textsuperscript{ 36,37,38,39,41} \\
mGTE-dense & Docs. (L3) & 41 & 0.834\textsuperscript{ 39,40,42} \\
BM25 & Docs. (G4) & 42 & 0.832\textsuperscript{ 41,43} \\
BGE-EN-ICL & Docs. (L3) & 43 & 0.831\textsuperscript{ 42} \\
mE5 & Docs. (G4) & 44 & 0.823 \\
mGTE-dense & None & 45 & 0.820\textsuperscript{ 46} \\
mGTE-dense & Docs. (Op) & 46 & 0.820\textsuperscript{ 45} \\
mE5 & Docs. (L3) & 47 & 0.817\textsuperscript{ 48,49} \\
Lajavaness & Docs. (Op) & 48 & 0.817\textsuperscript{ 47,49} \\
BM25 & Docs. (L3) & 49 & 0.815\textsuperscript{ 47,48,50} \\
mGTE-dense & Query (G4) & 50 & 0.813\textsuperscript{ 49} \\
BGE-EN-ICL & Query (G4) & 51 & 0.810 \\
BGE-M3-sparse & Gold & 52 & 0.807\textsuperscript{ 53} \\
mGTE-sparse & Gold & 53 & 0.805\textsuperscript{ 52} \\
Croissant & Query (G4) & 54 & 0.794\textsuperscript{ 55,56} \\
Croissant & Gold & 55 & 0.794\textsuperscript{ 54,56} \\
Croissant & None & 56 & 0.793\textsuperscript{ 54,55} \\
mE5 & Docs. (Op) & 57 & 0.790\textsuperscript{ 58} \\
BM25 & Docs. (Op) & 58 & 0.789\textsuperscript{ 57} \\
mE5 & None & 59 & 0.784 \\
Croissant & Docs. (G4) & 60 & 0.781\textsuperscript{ 61} \\
BGE-M3-sparse & Query (G4) & 61 & 0.781\textsuperscript{ 60} \\
BGE-M3-sparse & Docs. (G4) & 62 & 0.778 \\
Croissant & Docs. (L3) & 63 & 0.773\textsuperscript{ 64} \\
mGTE-sparse & Docs. (G4) & 64 & 0.771\textsuperscript{ 63} \\
mGTE-sparse & Query (G4) & 65 & 0.763\textsuperscript{ 66,67} \\
BGE-M3-sparse & Docs. (L3) & 66 & 0.763\textsuperscript{ 65,67,68} \\
mGTE-sparse & Docs. (L3) & 67 & 0.760\textsuperscript{ 65,66,68,69} \\
BGE-EN-ICL & Docs. (Op) & 68 & 0.760\textsuperscript{ 66,67,69} \\
mE5 & Query (G4) & 69 & 0.760\textsuperscript{ 67,68,70} \\
PSQ & None & 70 & 0.757\textsuperscript{ 69} \\
SPLADE++ & Query (G4) & 71 & 0.753 \\
Croissant & Docs. (Op) & 72 & 0.748 \\
BGE-M3-sparse & Docs. (Op) & 73 & 0.738\textsuperscript{ 74} \\
mGTE-sparse & Docs. (Op) & 74 & 0.737\textsuperscript{ 73} \\
BM25 & Query (G4) & 75 & 0.706 \\
SPLADE++ & None & 76 & 0.604 \\
BGE-M3-sparse & None & 77 & 0.449 \\
mGTE-sparse & None & 78 & 0.443 \\
BM25 & None & 79 & 0.417 \\
\bottomrule
\end{tabular}

  \end{minipage}
  \end{adjustbox}
  \caption{
    95\% bootstrap confidence intervals for Recall@100, using 1000 resamples.
    Numbers in subscripts indicate the 95\% interval of the system of the row overlaps with the interval of the systems in the subscripts
    \\G4: GPT-4o-mini. L3: Llama-3.2. Op: OpusMT.
  }
  \label{tab:ci_recall}
\end{table*}

\end{document}